\documentclass[aps,prl,preprint,groupedaddress,amsmath,amssymb,showpacs]{revtex4}

\usepackage{bm}
\usepackage{graphicx}

\usepackage[dvips]{color}

\begin{document}

\title{Crossover from normal to anomalous diffusion in field-aligned dipolar systems}

\author{Jelena Jordanovic and Sabine~H.~L.~Klapp}
\affiliation{Institute of Theoretical Physics, Sekr.~EW 7--1,
Technical University Berlin, Hardenbergstrasse 36,
             D-10623 Berlin, Germany}
\date{\today}

\begin{abstract}
Using molecular dynamics simulations we investigate the translational dynamics of particles with dipolar interactions in homogenous external fields. For a broad range of concentrations, we find that the anisotropic, yet normal diffusive behavior characterizing weakly coupled systems becomes anomalous both parallel and perpendicular to the field at sufficiently high dipolar coupling and field 
strength. After the ballistic regime, chain formation first yields
cage-like motion in all directions, followed by transient, mixed diffusive-superdiffusive behavior resulting from cooperative motion of the chains. 
The enhanced dynamics disappears only at higher densities close to crystallization.
\end{abstract}
\pacs{???}
\maketitle
There is a strong current interest in colloidal soft-matter systems displaying anomalous dynamic behavior 
beyond simple diffusion. A paradigm example are colloidal hard spheres which exhibit
a (repulsive) glass transition in near-perfect agreement with Mode Coupling theory \cite{Pusey87}.
Recent attention focuses, on one hand, on the anomalous {\em slow} dynamics in a broad variety of systems such as low-density colloidal gels \cite{Lu08}, cells \cite{Lebold09},
driven granular fluids \cite{Fiege09}, and disordered solids \cite{Kurzidim09}. In many cases the reduced mobility is accompanied by dynamic heterogeneities and cooperative motion as recently shown, e.g., in colloids \cite{Kob97,Chaud07} and in liquid crystals \cite{Patti09}. On the other hand, there is increasing interest in systems exhibiting {\em enhanced} mobility such as  rods between obstacles \cite{Hoefling08} and activated
systems such as catalytic nanorods \cite{Dhar06} or self-propelled bacteria \cite{Golestian09}.

In this letter we discuss the translational dynamics of {\em dipolar} systems which, as we will demonstrate, can exhibit {\em both}, slow and enhanced mobility. Prominent examples of dipole-coupled systems are suspensions of magnetic nanoparticles (ferrocolloids), polarizable colloids \cite{Leu09} or 
colloidal ''molecules'' (e.g. dumbells) with partial charges \cite{Mano03}, but also electric field-driven living cells \cite{Gupta09}. The strong current attention received by dipolar  systems (or, more generally,
systems with directional interactions) is born by their importance for self-assembly processes towards novel classes of equilibrium and non-equilibrium materials \cite{Gangwal10,Zhang07}. In zero field, the anisotropic interactions between permanent dipoles
can induce network formation \cite{Tlusty00} including the possibility of an equilibrium gel \cite{Blaak07}. On the contrary, the dynamics in an external field appears far less understood, although this situation is clearly most relevant for many practical applications of dipolar systems \cite{Odenbach}. Previous experimental and theoretical studies 
indicate that the single-particle motion under field can be strongly anisotropic \cite{Ilg05} and even sub-diffusive (in lateral directions) due to "caging" of the particles in the chains formed at sufficiently high coupling strength \cite{Furst00,Toussaint04,Belovs06}. However, only recently dynamic light-scattering experiments 
\cite{Mertelj09} reveal, in addition, strong evidence for superdiffusive motion and dynamic heterogeneity. Motivated by these exciting and seemingly somewhat contradictory findings, we have performed extensive molecular dynamics (MD) simulations of a simple dipolar model system. Analyzing separately the microscopic motion parallel and perpendicular to the field
for a broad range of parameters we find, for the first time, a coupling strength-induced crossover from normal to anomalous dynamics. The latter is characterized by a
step-wise behavior involving sub-diffusion, directed motion and dynamic heterogeneities. 

Our model fluid consists of dipolar soft spheres (DSS) of diameter $\sigma$ with embedded, permanent point dipole moments $\bm{\mu}_i$.
The total pair interaction at distance $r=|\mathbf{r}|$ consists of a (truncated and shifted) soft-sphere repulsion,
$u_{\text{SS}}(r)=4\epsilon\left(\sigma/r\right)^{12}$,
and the long-range, anisotropic
dipole-dipole potential, $u_{\text{DD}}(12)=r^{-3}
\left[\bm{\mu}_1\cdot\bm{\mu}_2-3\left(\bm{\mu}_1\cdot\mathbf{r}\right)
 \left(\bm{\mu}_2\cdot\mathbf{r}\right)/r^2\right]$. We employ constant-temperature MD simulations with
 $N=500-1372$ particles. Thus, we focus on the direct (conservative) effects on the dynamics rather than on those stemming from a solvent, in accordance with 
 other recent studies (see, e.g., \cite{Blaak07}). The dipolar interactions are treated using the Ewald method with conducting boundaries (see \cite{Jordanov09} for further details).
 Simulations are carried out at a reduced density $\rho^{*}=\rho\sigma^3=0.05$ (which is typical for the experiments in \cite{Mertelj09}) and a
 reduced temperature $T^{*}=k_{\text{B}}T/\epsilon=1.35$ 
 (with $k_{\text{B}}$ and $T$ being Boltzmann's constant and true temperature, respectively).
 The dipole moment is chosen such that the resulting coupling
 parameter $\lambda=\mu^2/k_{\text{B}}T\sigma^3$ has values $1-7$, consistent with the
 range considered in recent ferrofluid experiments \cite{Mertelj09,klokkenburg06}. The fluids are subject to 
a homogeneous magnetic field $\mathbf{H}=H\mathbf{e}_{\text{z}}$ with fields strengths
up to the (still realistic \cite{klokkenburg06}) value 
$H^{*}=\mu H/k_{\text{B}}T\approx 100$. 

\begin{figure}
\centering
\includegraphics[width=8cm,clip]{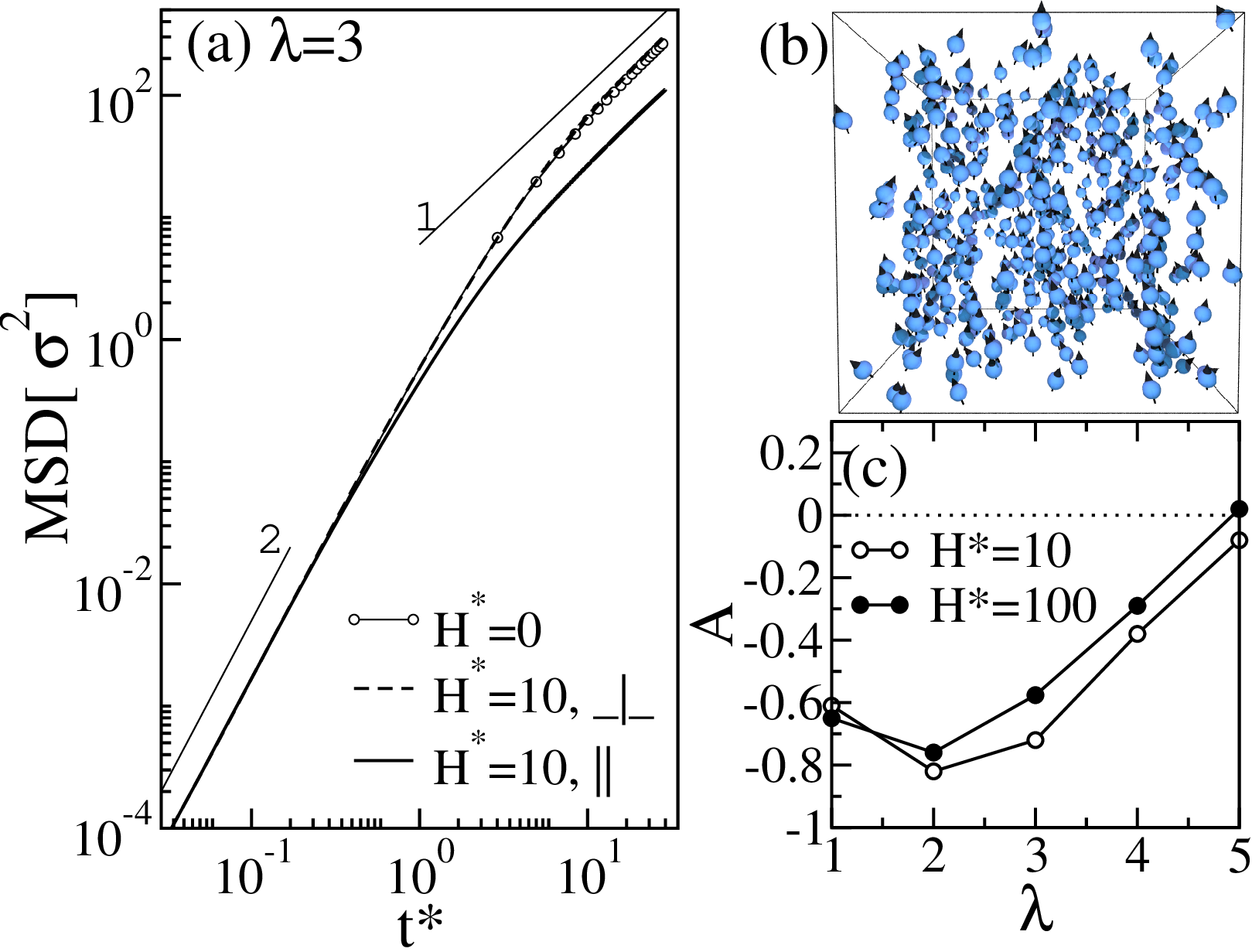}
\caption{(Color online) a) MSDs for $\lambda=3$ at $H^{*}=0$ and $10$;  b) simulation snapshot at $\lambda=3$ and $H^{*}=10$;
c) anisotropy ratio as function of dipolar coupling strength.}
\label{lambda_small}
\end{figure}

One key quantity of our analysis is the translational mean-squared displacement (MSD), $\Delta \mathbf{r}^2(t)=\langle\left(\mathbf{r}(t)-\mathbf{r}(0)
\right)^2\rangle$. 
Representative MSDs  for a moderate coupling parameter ($\lambda=3$) are shown in Fig.~\ref{lambda_small}a). 
For $H>0$ we differ between the MSDs parallel and perpendicular to the field
direction. This is sensible due to the structural anisotropy resulting from the formation of elongated clusters
along the field, see Fig.~\ref{lambda_small}b). Despite these (small) clusters, all of systems have "normal" translational dynamics in the sense that there is an initial ballistic regime, 
$\Delta \mathbf{r}_{\parallel(\perp)}^2(t)\propto t^2$, after which the MSD directly goes over into a diffusive regime $\Delta \mathbf{r}_{\parallel(\perp)}^2(t)\propto t$. 
The latter, Einstein-like behavior allows us to define diffusion constants $D_0$ (for $H=0$) and $D_{\parallel}$, $D_{\perp}$ ($H>0$) in the standard way. Of particular interest is the anisotropy ratio $A=\left(D_{\parallel}-D_{\perp}\right)/D_0$, results for which are plotted in Fig.~\ref{lambda_small}c). At the coupling strengths considered, motion perpendicular to $\mathbf{H}$ is ''faster" than along the field, i.e., $A<0$ (consistent with earlier simulation
results \cite{Ilg05}), despite the fact that the clusters form in $z$-direction. Thus
the lateral translational fluctuations of the particles are still so large that their diffusive behavior resembles disks than rods (where one expects $A>0$). 
However, as Fig.~\ref{lambda_small}c) clearly indicates, there is
a crossover towards different dynamic behavior as $\lambda$ increases beyond $\approx 5$.

\begin{figure}
\centering
\includegraphics[width=8cm,clip]{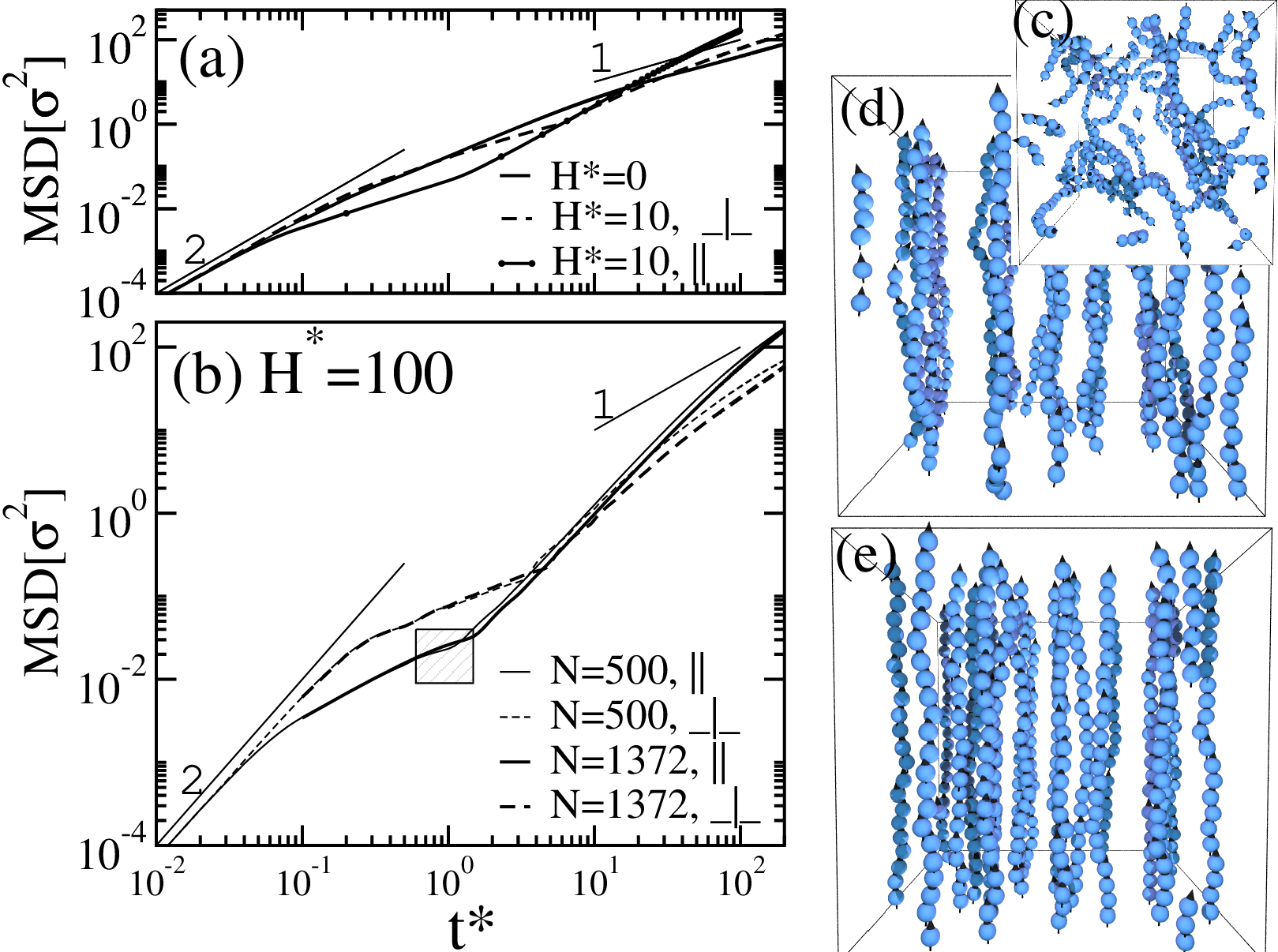}
\caption{(Color online) a)-b) 
MSDs for $\lambda=7$ and $H^{*}=0$, $10$, $100$. b) includes data for two system sizes; the box indicates the sub-diffusive regime.
c)-e) corresponding snapshots.}
\label{msd_7}
\end{figure}

\begin{figure}
\centering
\includegraphics[width=8cm,clip]{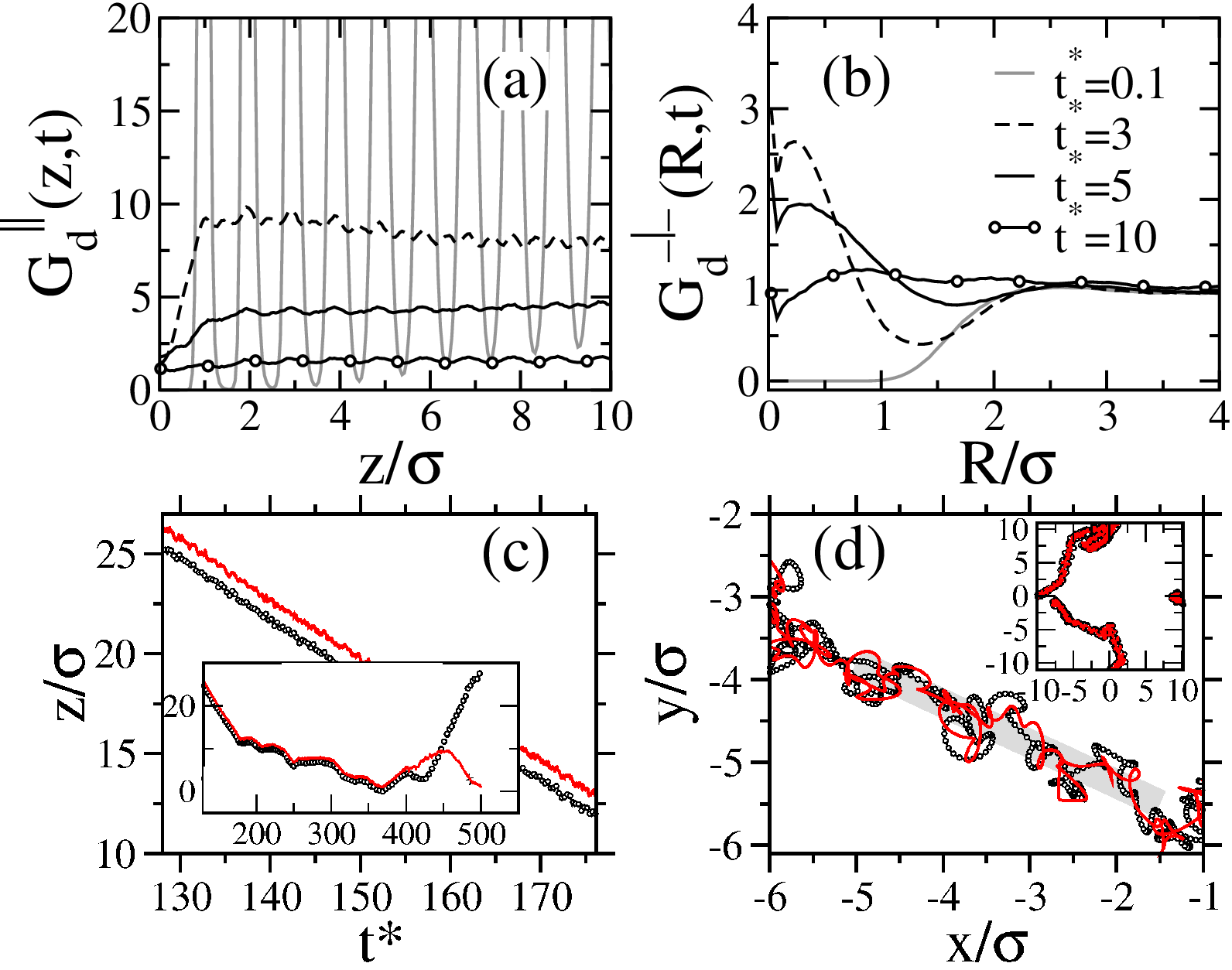}
\caption{(Color online) a), b): Distinct-van Hove functions ($\lambda=7$, 
$H^{*}=100$). c), d): Trajectories of two initially neighboring particles (insets show a larger time scale).}
\label{correlations}
\end{figure}

Typical MSD's for this strongly coupled regime are plotted in Fig.~\ref{msd_7} along with corresponding simulation snapshots ($\lambda=7$).
At zero field, the particles arrange into a globally isotropic network of worm-like chains [Fig.~\ref{msd_7}c)], as it is typical for dilute, strongly interacting dipolar systems \cite{Tlusty00,Blaak07}.
The corresponding MSD still behaves "normal" in the sense described above, consistent with MD results for dipolar dumbbells \cite{Blaak07} at comparable
values of $\lambda$. With field, long chains form which become more and more straight upon increase of $H$. 
We first consider the case $H^{*}=100$ [see Figs.~\ref{msd_7}b) and e)].
After the initial ballistic regime, the motion
of a particle significantly slows down due to the confinement by its nearest neighbors, yielding a sub-diffusive regime in the MSD characterized by a time dependence $t^{\gamma}$ with 
$\gamma<1$. This holds not only in $z$-direction (where sub-diffusion occurs for $0.6\lesssim t^{*}\lesssim 1$ with
$\gamma_{\parallel}\approx 0.61$), but also perpendicular to $\mathbf{H}$
($0.3\lesssim t^{*}\lesssim 3.3$, $\gamma_{\perp}\approx 0.69$), indicating pronounced spatial-temporal correlations in all spatial directions. We characterize
these correlations by the distinct parts of the van Hove correlation functions $G_{\text{d}}^{\parallel}(z,t)$ and $G_{\text{d}}^{\perp}(R,t)$ with $z$ and $R$ being the
longitudinal and lateral distance, respectively. Results are plotted in Figs.~\ref{correlations}a) and b). 
For $t\rightarrow 0$ the functions $G_{\text{d}}^{\parallel(\perp)}$ coincide with the corresponding static pair distribution functions.
The data for $G_{\text{d}}^{\parallel}$ at $t>0$ clearly reflect long-range correlations,  and thus, a "trapping" of particles along the chain, which starts to die out only at 
times {\em beyond} the sub-diffusive regime of the MSD (i.e., $t^{*}\gtrsim 1.1$).
The exponent $\gamma_{\parallel}\approx 0.61$ within the sub-diffusive regime is close
to the value of $0.63$ found 
in the MSD of interior beads in linear polymer chains \cite{Baschnagel00} prior to crossover to free diffusion. In perpendicular directions [see Fig.~\ref{correlations}b)], 
the correlation "hole" 
at $t\rightarrow 0$ reflects that positions directly besides a particle are disfavored due to the repulsive dipolar interactions associated to this configuration. At later times, however,
we observe development of a peak at $R\rightarrow 0$, indicating that the original position of a particle is replaced by another particle. These strong and long-living 
correlations explain the sub-diffusive behavior of the perpendicular MSD, characterized
by an exponent $\gamma_{\perp}=0.69$. A similar exponent ($\gamma\approx 0.75$)
has been found in experiments probing lateral diffusion in magnetorheological systems \cite{Furst00,Toussaint04} involving chains of {\em induced} (and thus, perfectly aligned)
dipoles. This similarity already indicates that the strong suppression of orientational fluctuations, i.e., a strong field, is an important ingredient
for the observed sub-diffusive behavior in our present system consisting of permanent dipoles.
Indeed, the MSD's at the somewhat weaker field $H^{*}=10$ [see Fig.~\ref{msd_7}a),d)] do not reveal a sub-diffusive regime. Instead,
one merely observes a slowing down ($\gamma_{\parallel}=1.13$, $\gamma_{\perp}=1.12$) as compared to the initial ballistic regime.

\begin{figure}
\centering
\includegraphics[width=8cm,clip]{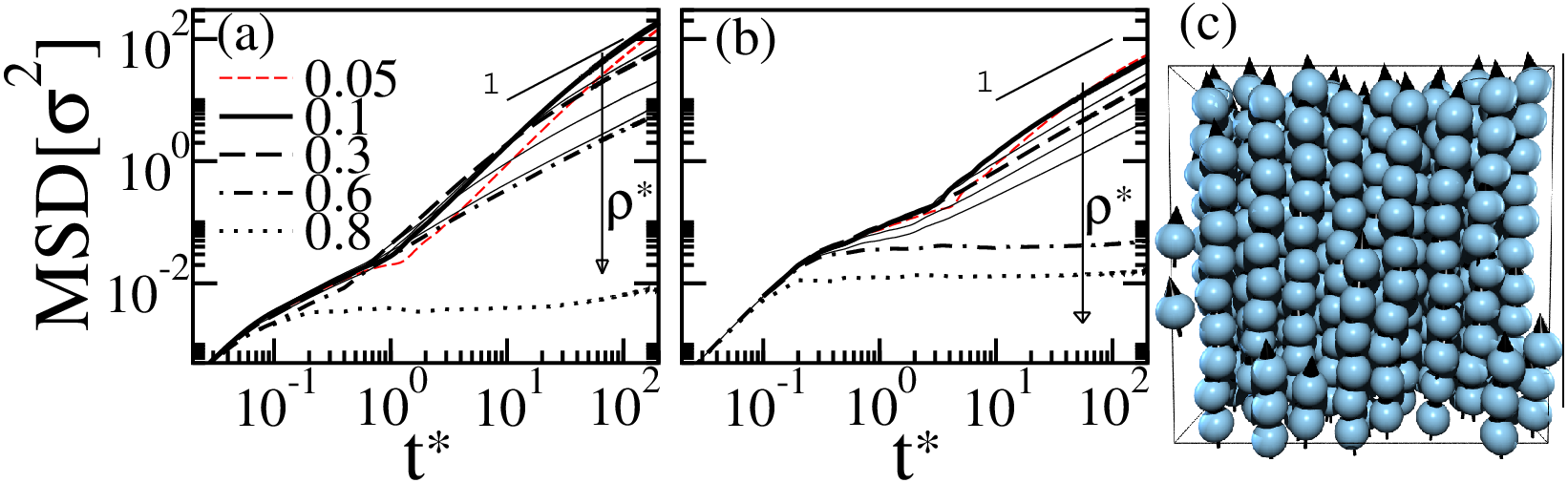}
\caption{(Color online) MSDs a) parallel b) perpendicular to $\mathbf{H}$ at various densities $\rho^{*}$ ($\lambda=7$, $H^{*}=100$, $N=864$). Part c) is a snapshot
at $\rho^{*}=0.8$}
\label{msd_dense}
\end{figure}

However, both anisotropic systems ($H^{*}=100$ and $H^{*}=10$) display, at $\lambda=7$, a further intriguing feature not present in weakly coupled systems considered
in Fig.~\ref{lambda_small}. Specifically, we observe in Figs.~\ref{msd_7}a)-b)
a ''kink'' in the MSDs, which separate the intermediate time range discussed above and a later time regime where the MSDs grow {\em faster} than linearly in time. This 
behavior strongly contrasts that in other disordered systems (see, e.g., \cite{Baschnagel00}), where the sub-diffusive regime, if present,
changes directly into true diffusion ($\gamma=1$). 
In our present system, the initial exponent within the ''fast'' regime (after the kink) lies between $1.5$ and  $1.8$; however, these values decrease in time
and also strongly depend on $H^{*}$, as does the time related to the kink. 
We conclude that there is a (transient) superposition of diffusive and ballistic ($\gamma=2$) behavior, with 
a tendency towards normal diffusion (visible particularly in the lateral part) at large times. In fact, at the low density considered  ($\rho^{*}=0.05$) we did not reach the true diffusive regime particularly for the longitudinal part in the total simulation time (contrary to what is found at larger $\rho$, see Fig.~\ref{msd_dense}). 
To test possible finite-size effects we have repeated the simulations with $N=1372$ particles, the changes [see Fig.~\ref{msd_7}b)], however, being only marginal. 

A direct illustration of the superposition of diffusive and ballistic motion is provided by typical trajectories plotted in the bottom of Fig.~\ref{correlations}. 
In $z$-direction [\ref{correlations}c)], one clearly recognizes 
small random-walk-like movements of the particles around a mean path consisting of essentially {\em straight} stretches, the latter giving rise to the "fast" regime of the parallel MSD 
[see Fig.~\ref{msd_7}]. Similar random movement around a mean path occurs perpendicular to $\mathbf{H}$ [Fig.~\ref{correlations}d)], where, 
however, the stretches are shorter and, consequently, the growth of the MSD (after the kink) is somewhat less pronounced. Moreover, 
in both directions, the motion of two particles
which are nearest neighbors at $t=0$ is correlated over an extremely long time, consistent with the behavior of $G_{\text{d}}^{\parallel}$ (and $G_{\text{d}}^{\perp}$)
shown in the top of Fig.~\ref{correlations}. These observations suggest that, within the fast regime, the particles move collectively as chains (indeed,
at the large field considered, there are essentially no unbounded particles). Moreover, the stretches in the trajectories indicate that the entire chains can 
move quasi-independently (i.e., ballistically) over a finite time interval before they are slowed down due to interactions with particles in other chains.

\begin{figure}
\centering
\includegraphics[width=8cm,clip]{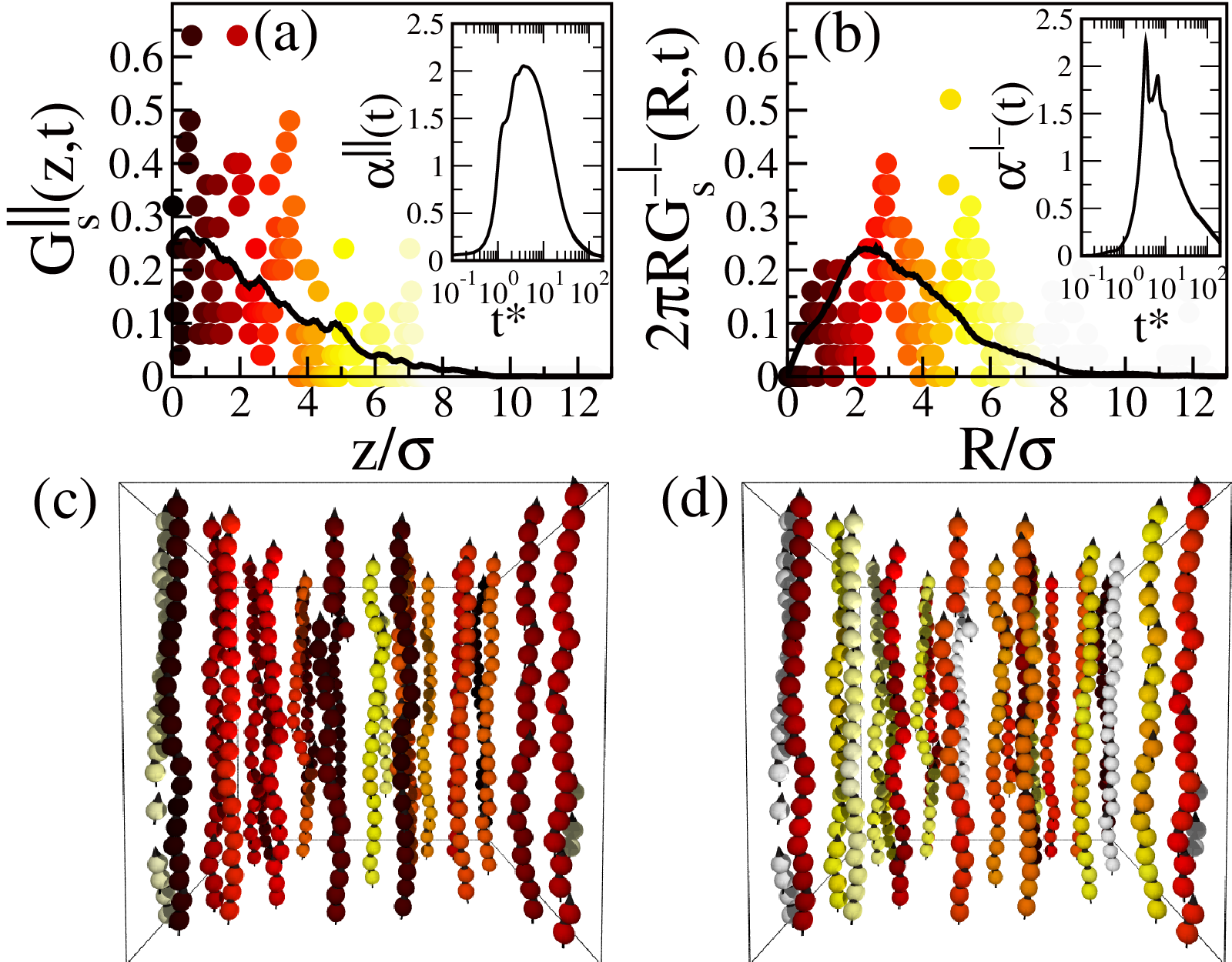}
\caption{(Color online) a), b) The functions $G_{\text{s}}^{\parallel(\perp)}$ at $t^{*}=20$ with corresponding instantaneous data. The insets show 
the non-Gaussian parameters $\alpha_{\parallel(\perp)}(t)$.
c), d) Snapshots with particles colored according to the code in a) and b).}
\label{Gs_snap}
\end{figure}

The occurrence of cooperative motion of the particles as chains is further confirmed by the behavior of the self parts 
of the van Hove functions, $G_{\text{s}}^{\parallel}(z,t)$ and $G_{\text{s}}^{\perp}(R,t)$ measuring the probability that a particle moves over a certain distance in a time $t$.
Exemplary results for time $t^{*}=20$ (i.e., within the ''fast'' regime) 
are shown in Fig.~\ref{Gs_snap}. 
Particularly the parallel part strongly deviates from the usual Gaussian shape characterizing diffusive systems. Similar deviations occur at other times (and in the perpendicular part),
yielding large values of the ''non-Gaussian"-parameter 
$\alpha(t)=\langle \left(\Delta \mathbf{r}\right)^4\rangle/\langle\left(\Delta\mathbf{r}\right)^2\rangle^2-1$
(see insets in Fig.~\ref{Gs_snap}). At $t^{*}=20$, specifically, the buckled structure of $G_{\text{s}}^{\parallel}$ [Fig.~\ref{Gs_snap}a)]  suggests
that the system consists of several portions (''populations'') of particles. To better identify these populations 
we have resolved the instantaneous data for $G_{\text{s}}^{\parallel(\perp)}$ with respect to the actual particles in the system as illustrated
in Fig.~\ref{Gs_snap}c),d). One clearly recognizes that particles related to one ''population'' in 
$G_{\text{s}}^{\parallel}$ belong to the {\em same} chain in the actual configuration. 
 %
%
Also, comparison of Figs.~\ref{Gs_snap}c) and d) shows that chains which are fast in $z$-direction move more slowly
in perpendicular directions and vice versa. 

Taken altogether, we interpret the ''kink'' in the MSDs (and thus, the superdiffusive behavior), as a consequence of the motion of the entire chains, which
over certain times, interact only weakly with particles of other chains. This interpretation prompts the question what would happen at higher densities, where chain-chain interactions should be more prominent. 
To this end, we plot in Fig.~\ref{msd_dense} results for MSD's in the range $0.05\leq \rho^{*}\leq 0.8$. The simulations have been performed at
the same strong coupling conditions as before, yielding essentially percolated chains in $z$-direction.
An inspection of the data in Fig.~\ref{msd_dense} reveals that the ''kink'' indicating onset of (transient) superdiffusive motion along $\mathbf{H}$
survives up to $\rho^{*}\simeq 0.3$ in field direction; at higher $\rho^{*}$ one observes normal dynamics where the initial ballistic regime directly changes 
into diffusion. Perpendicular to $\mathbf{H}$, the disappearance of the kink occurs already $\rho^{*}\simeq 0.1$; at $\rho^{*}\gtrsim 0.3$ 
plateau-like regions in the MSD appear. Indeed, at high densities 
one expects \cite{Tao91} formation of body-centered-tetragonal structures with neighboring chains being out of registry, a tendency which is visible in the
snapshot in Fig.~\ref{msd_dense}c). Clearly, this lateral ordering induces additional dynamic effects not present at small $\rho^{*}$.

To summarize, our MD simulation results demonstrate that chain formation of
strongly coupled, dilute or moderately dense dipolar fluids in external fields gives rise to anomalous translational dynamics. The latter involves {\em both},
sub-diffusive behavior at intermediate times and mixed diffusive-superdiffusive behavior at later times, consistent with recent ferrofluid experiments \cite{Mertelj09}.
According to our analysis
in the superdiffusive regime, the particles move collectively as chains, a behavior which appears as dynamic heterogeneity on the single-particle level. 
From a fundamental point of view, a particularly intriguing result of our study is that the systems may display enhanced dynamics already {\em in equilibrium}, contrary to {\em active} 
(bio-) materials such as self-propelled bacteria \cite{Golestian09} and catalytic nanorods \cite{Dhar06}, which are intrinsically out of equilibrium.
From an application perspective, we expect our results to be relevant not only for ferrofluids, but for the much broader class of colloidal systems 
where external fields induce {\em directed self-assembly}, examples being patchy colloids such as metallodielectric (Janus) particles \cite{Gangwal10} and living cells
suspensions \cite{Gupta09}.
Clearly more detailed experiments are needed to confirm our findings, such as the "kink" separating the dynamic regimes in the MSD. On the theoretical side,
points deserving further attention are the role of orientational fluctuations at moderate fields, as well as solvent (hydrodynamic) effects
which are so far neglected. Work in these directions is in progress.
\begin{acknowledgments}
Financial support from the German~Science~Foundation via grant~KL1215/6 is gratefully acknowledged.
\end{acknowledgments}

\end{document}